\documentclass[useAMS,usenatbib]{mn2e} 
\usepackage{epsfig} 
\usepackage{url} 
\usepackage{color}

\def\kms{\rm{ \, km \, s^{-1}}} 
\def\sm{\rm{ \, M_\odot}} 
\def\gy{\rm{ \, Gyrs}} 
\def\my{\rm{ \, Myrs}} 
\def\vc{V_{circ}} 
\def\mpc{\rm{ \, Mpc}} 
\def\kpc{\rm{ \, kpc}}

\title [ Formation of Galactic Pseudo-bulges via Gas Rich Major Mergers] { Pseudo-bulge formation via major mergers} 
\author [J. A. Keselman \& A. Nusser] { J. A. Keselman$^{1}$\thanks{E-mail: kari@tx.technion.ac.il}, and A. Nusser$^{1,2}$\\
$^{1}$Physics department, Technion, Haifa 32000, Israel\\
$^{2}$Asher Space Research Institute, Technion, Haifa 32000, Israel }

\begin{document} 
\pagerange{ 
\pageref{firstpage}-- 
\pageref{lastpage}} 
\maketitle \label{firstpage} 
\begin{abstract}
It is widely accepted that within the framework of $\Lambda$CDM a significant fraction of giant-disk galaxies has recently experienced a violent galactic merger. We present numerical simulations of such major  mergers of gas-rich pure disk galaxies, and focus on the innermost stellar component (bulge) of the disk remnants. The simulations have high spatial and mass resolutions, and resolve regions deep enough to allow bulge classification according to standard kinematical and structural characteristics. In agreement with recent studies we find that these bulges are dominated by stars formed in the final coalescence process. In contrast to the common interpretation of such components as classical bulges (i.e. similar to intermediate luminosity ellipticals), we find they are supported by highly coherent rotations and have S\`ersic indices $n<2$, a result leading to their classification as pseudo-bulges. Pseudo-bulge formation by gas rich major mergers of pure disks is a novel mode of pseudo-bulge formation; It complements pseudo-bulge growth by secular evolution, and it could help explain the high fractions of classically bulge-less giant disk galaxies, and pseudo-bulges found in giant Sc galaxies. 
\end{abstract}

\section{Introduction} A galactic bulge in a spiral galaxy is morphologically defined as the excess of light in the central region of the galactic disk. With this simple definition, recent numerical simulations which include sub-grid Inter-Stellar Medium (ISM) physics are beginning to successfully reproduce the observed bulge masses \citep[e.g.][]{steinmetz02,parry09,hopkins10}. However, bulges have complex dynamical properties that are still inaccessible by means of current simulations. Detailed observations of nearby galaxies reveal spiral galaxies harbouring classical bulges which are supported by random motions of their stellar components. Others spirals, however, show evidence for pseudo-bulges, those are bulges that are mainly supported by coherent rotation. Current models for bulge formation remain challenged when confronted with the observed division into classical and pseudo-bulges. The key issue is that there are fewer classical bulges than expected from models for bulge formation by major mergers in the context of the standard hierarchical $\Lambda CDM$ paradigm \citep[see e.g.][]{white78,cole94,peebles93}. In these models major merging are ubiquitous and are thought to produce remnants that are centrally dominated by classical bulges. \cite{kormendy10} studied galaxies with $\vc>150\kms$ within a sphere of radius $8\mpc$ centered on our Galaxy. They find that 11 out of 19 galaxies show no sign of classical bulges. \cite{fisher11}, using Spitzer $3.6\mu m$ detector and Hubble space telescope data, find that within an $11\mpc$ sphere the dominant galaxy type has pure disk characteristics, whether counting by number, star formation rate, or stellar mass.

The classification into classical and pseudo-bulges could be made on the basis of a variety of observational characteristics \citep{kormendy04,athana05}. In addition to kinematics, of particular importance is the distribution of stars in the galactic components. The characterisation of the stellar distribution is assesed by the S\`ersic index $n$ \citep{sersic63,sersic68}, used in the following parametrization 
\begin{equation}
	\label{eq:sersic} \Sigma(r)=I_e \exp \left\lbrace -b_n \left[ \left( \frac{R}{R_e} \right)^{1/n}-1 \right] \right\rbrace, 
\end{equation}
of the the observed (surface) stellar intensity profile, $\Sigma(r)$. This form includes two free parameters, $R_e$ is an effective radius enclosing half the modelled mass, and $I_e$ which is the intensity at $R_e$. The parameter $b_n$ is defined by the latter requirement, i.e. it is calculated from $I_e$ and $R_e$. The stellar distribution in both bulges and pseudo-bulges is well approximated by the form \ref{eq:sersic}, typically with $n>3$ for classical bulges, and $n<2$ for pseudo-bulges, closer to exponential disks ($n=1$).

Here we explore the possibility that violent major merging could actually lead to the formation of pseudo-bulges. We find that gas rich major mergers of pure disk galaxies could indeed produce inner cores with morphological and kinematical properties of the observed pseudo-bulges. This does not mean that violent mergers will be the main mode of pseudo-bulge formation since it requires the existence of a significant population of pure disk galaxies. However, it could greatly mitigate the tension between standard bulge formation models and observations. The physical conditions for pseudo-bulge formation by secular growth during the quiescent later phase of galaxy evolution, remain more likely to be satisfied.

Like most recent studies of galactic bulge formation via mergers, we take into account key physical processes such as radiative dissipation, star formation, and feedback. The reason these are needed is that classical bulges obtained from dry major mergers do not resemble intermediate luminosity ellipticals like the observed bulges do, i.e. flattening by rotation \citep{davis83}, S\`ersic parameter $n$ that scales with luminosity \citep{graham03}, color-magnitude relations \citep{balcells94}, and metallicity-luminosity relations \citep{jablonka96}. Bulges created by dry mergers are generally more similar to giant ellipticals or diffuse stellar halos. This can be seen, for e.g., in Fig. 5 in \cite{cox06}, where remnants of gas-less galaxies show no sign of a significant inner component. 

We now review a few recent studies of galactic bulge formation via mergers, and focus on difficulties regarding the classification into classical versus pseudo-bulges. \cite{cox06} study remnants of galactic mergers and analyse them using observational methods, while treating the whole remnant as a single component. The whole remnant being a single component, they find large effective radii with mean of $4\kpc$, a number significantly larger than measured in this work. Also, their simulations do not have the required spatial resolution to measure the S\`ersic index of the inner component. \cite{hopkins10} uses simulations similar to \cite{cox06}. In their decomposition of the remnants into various components, bulges are assigned only those particles with weak rotational support. Essentially, any disk-like bulge is not treated, and is assumed to be part of the disk. \cite{robertson06} use similar simulations, but adopt two methods to decompose the remnants. First, much like what we do here, a remnant is decomposed into a bulge, thin disk, thick disk, and a spheroid, according to their stellar surface density. This decomposition is then used mainly to derive relative masses. Second, they decompose the remnant into old stars, that formed a few hundred $\my$ before the peak of the star formation rate (hereafter SFR), and young stars, formed a few hundred $\my$ afterwards. They find that young stars are highly rotationally supported (disk like) while old stars are not (classical bulge). The composition of the remnant core (whether it is composed mainly of young or old stars) is not addressed. The stellar population formed during the main SFR peak is not analysed as a separate component. However, as we show here, these stars may have a major contribution to the central density of the galaxy \citep[see also][]{cox06}, constituting a significant component which could be classified as a pseudo-, rather than a classical bulge. Pure bulge-dominated spheroids, with rotating central components formed via dissipative inflows and "in-situ" star formation by the inflowing gas in the final coalescence is very well-studied in previous work. See for e.g. \cite{mihos94,hopkins08b,hopkins08a,hopkins09a,hopkins09b,hoffman09,hoffman10,jesseit09}.

None of these studies directly address the question of whether pseudo-bulges could form via major mergers. Here, we revisit the problem of bulge formation by major mergers through simulations, including the usual essential processes of star formation, SN feedback and radiative dissipation of gas. Our simulations are of sufficiently high spatial and mass resolution to allow a realistic analysis of the simulated stellar cores in more details than done before. We find that despite their violence, dissipative major mergers can produce a central stellar component with highly coherent rotation.

The structure of the paper is as follows. In \S \ref{sec:genset} we give a high level description of the numerical experiment. In \S \ref{sec:numerical_model} we describe the numerical model: we start with a brief description of the simulation code, then in \S \ref{sec:galactic_model} we describe the galactic model used to create the simulated galaxies. In \S \ref{sec:merging} we describe the simulations (orbits, initial positions, integration time, and so on), in \S \ref{sec:remnants} the structural and kinematic properties of the simulated remnants, in \S \ref{sec:num_disc} numerical issues, and we conclude with a general discussion about the implications of our findings in \S \ref{sec:discussion}. We generate galaxies with massive stellar and gaseous disks in equilibrium using a new method, described in detail in appendix \ref{app:galaxy_generation}.

\section{The general set-up} \label{sec:genset} We wish to derive conditions under which merging of pure disk galaxies is likely to form pseudo-bulges at the cores of disk-like galaxies. We address this problem using a suite of merging simulations of initially isolated multi-component galaxies which include gas, stars and collisionless dark haloes. The wide range of possible configurations of the merging galaxies is ideally probed in a large scale cosmological simulation. Unfortunately such a simulation including all relevant physical effects, and the sub-$\rm kpc$ resolution required to model the cores of galaxies, is unavailable. Instead we aim at extracting robust general conclusions from a few detailed merging simulations. 
\begin{figure*}
	\centerline{\epsfig{figure=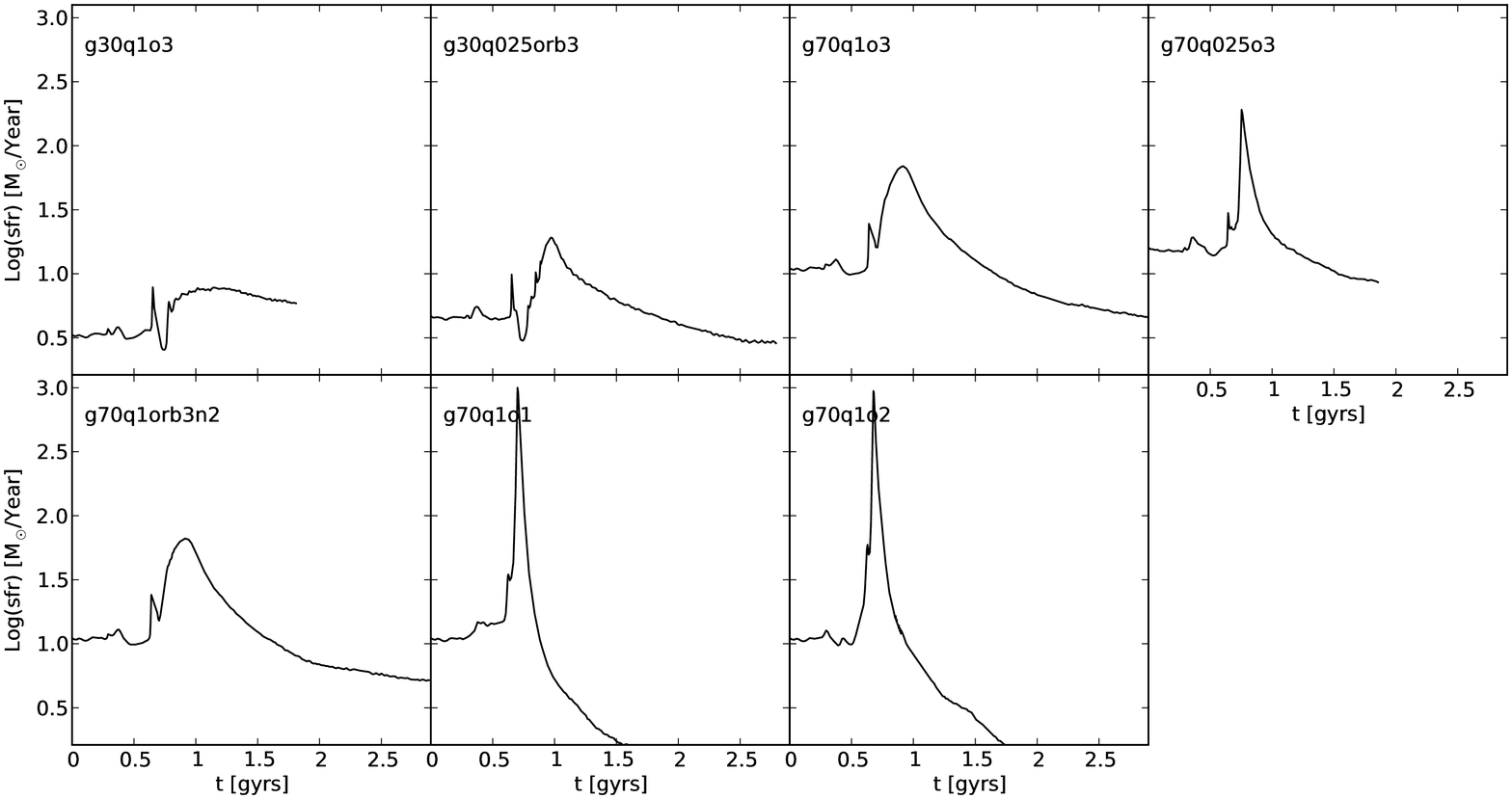, width=460pt, height=230pt}} \caption{The star formation rate as a function of time since the beginning of the simulation.} \label{fig:r_sfh} 
\end{figure*}
\begin{figure*}
	\centerline{\epsfig{figure=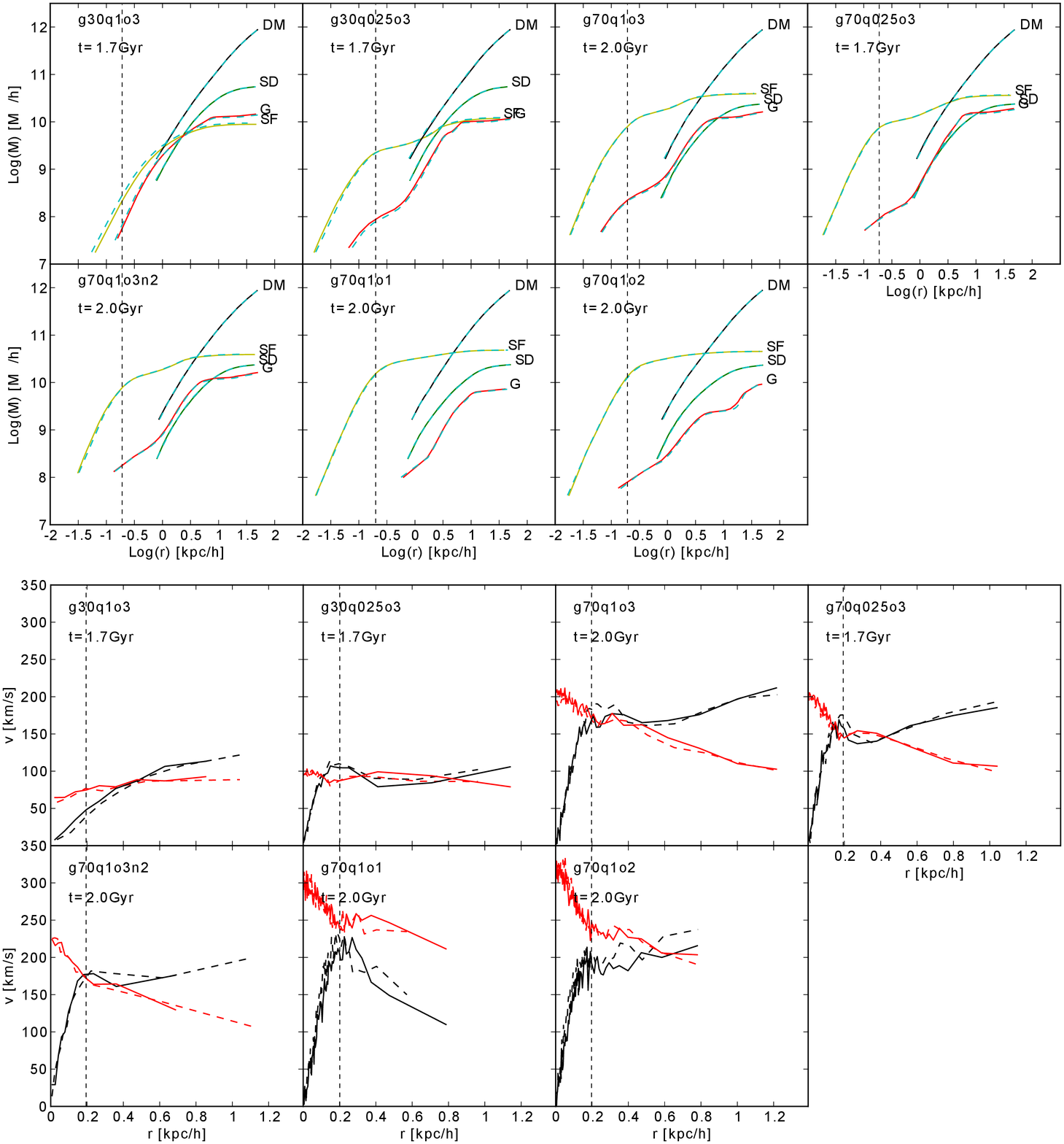, width=460pt, height=500pt}} \caption{ Equilibrium of the remnants in terms of mass profile and velocity moments. {\it Top panel}: Cumulative mass within radius $r$ from the center. Solid lines are measured at $t=t_{rf}$, and dashed lines are measured $143$ million years later. Here, G is for gas, SD for stellar disk, SF for stars formed during the collision, and DM for dark-matter. Vertical dashed lines denote formal numerical resolution. {\it Bottom panel}: Velocity moments of remnants. Velocities are measured like an observer along the edge-on line of sight. Black lines denote rotational velocity, and red lines central velocity dispersions. Vertical dashed lines denote formal numerical resolution. } \label{fig:r_eq} 
\end{figure*}

We perform seven simulations of mergers of identical pairs of pure disk galaxies. Prior to the merging, the original galaxies fall into four configurations corresponding to two choices for the gas mass fraction in the disk and two choices for the effective equation of state of the ISM. These original galaxies have been generated in isolation, all sharing the same choices for the masses in the baryonic and dark components, the halo density profile, and the disk scale lengths (vertical and radial). They also have the same number of baryonic and DM simulation particles, and the same criteria for setting the force softening lengths. The original galaxies all have $r_{200}=232\kpc$, roughly like the Milky-Way \citep{yin09}. As mentioned in \S \ref{sec:galactic_model} the DM distribution does not strictly follow an NFW profile, however, at small radii, the profile is close to an NFW form with $c_{_{NFW}}=9$. The values of all relevant parameters are listed in table \ref{tab:galaxies} and explained in details in the caption. The table also lists a fifth galaxy (g0) which is only used to test the validity of the scheme used in generating equilibrium configurations of collisionless stellar disk embedded in a spherical DM halo.

\section{Numerical model} \label{sec:numerical_model} We use a modified version of the Gadget2 hydrodynamical N-body code \citep{gadget2}, which includes radiative dissipation of gas energy, star formation and SN feedback, as described in \cite{springel03}. In this model, The ISM is cooled by line emissions of a primordial gaseous mix, and heated by SN energetic feedback. It can be shown that under some reasonable assumptions, in star-forming regions, the ISM is in a stable phase, and exhibits an effective equation of state (EEOS), in which the pressure is a function of density. This is the state of the ISM assumed in the simulations, in regions with gas density above a threshold $\rho_{th}$. Since this is a sub-grid model, and the feedback mechanism is not completely understood, there is some freedom in choosing the ISM pressurization. The parameter Q is used to linearly interpolate the EEOS between the full model and an isothermal EOS (Q=0 means there is no feedback, and $Q=1$ means the full model is used, with the chosen efficiency parameters). \cite{hopkins10b} compare various choices for the Q parameter to observations, with the conclusion that Q $\sim 0.1$-$0.3$ provide the best match. We use the same parameters as used by \cite{springel05}, where the efficiency of the evaporation process $A_0=4000$, the mass fraction of stars that are short-lived and die as supernovae $\beta=0.1$, the cold gas cloud temperature $T_{cloud}=1000{\rm K}$, the effective SN temperature $T_{SN}=4\times 10^8{\rm K}$, and the gas consumption time-scale $t_o^*=8.4\gy$. This choice gives a stellar formation rate of $\sim 1\sm$ per year for isolated galaxies with structure similar to the Milky-Way. In the numerical representation, each gaseous particle may spawn up to $n_s$ stellar particles of mass $m_g/n_s$ where $m_g$ is the initial mass of the gas particle. In order to assess the numerical effect of $n_s$ We use simulations both with $n_s=6$, and $n_s=2$. Throughout this study we adopt the Hubble parameter $h=0.7$.
\begin{figure}
	\centerline{\epsfig{figure=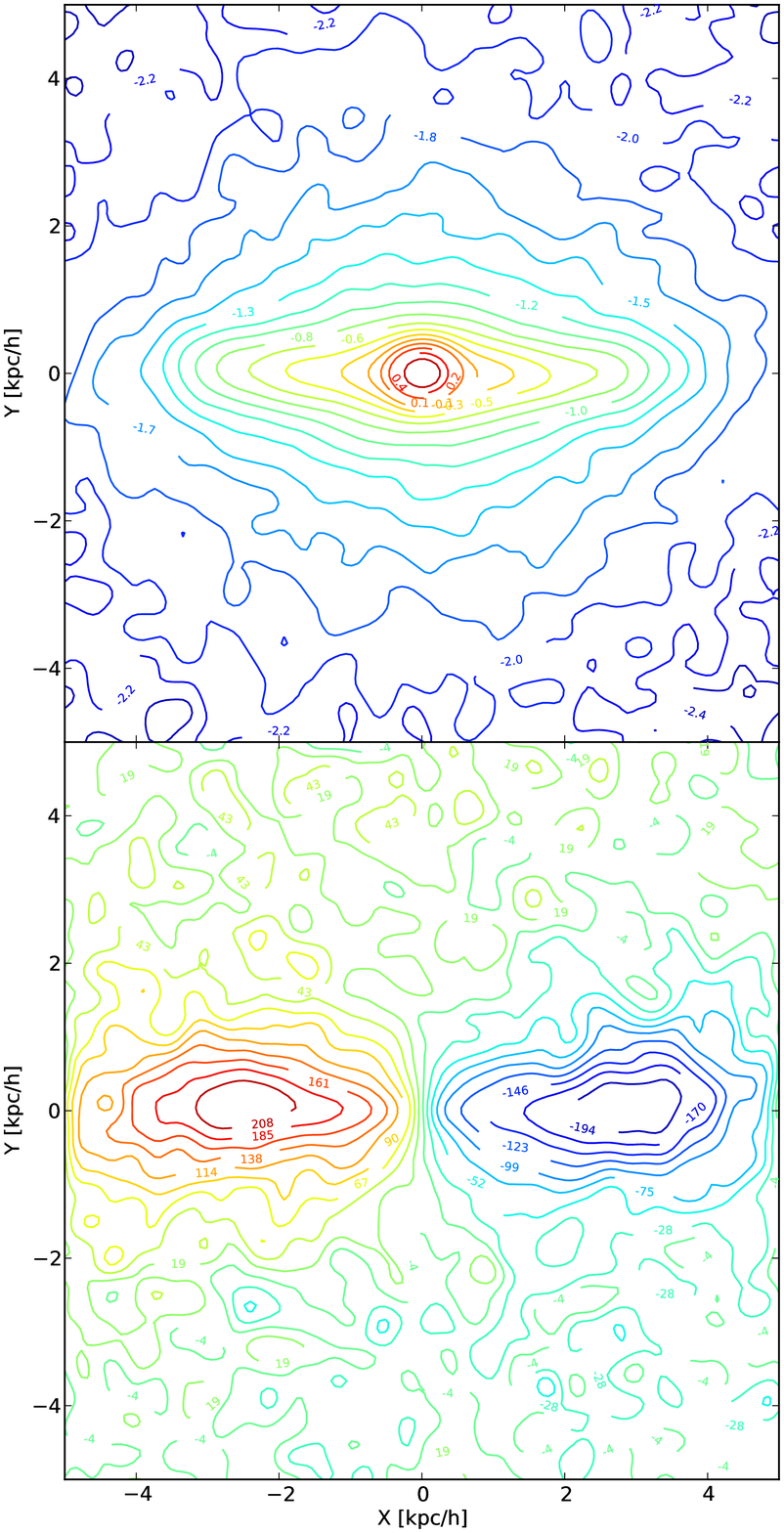, width=230pt, height=440pt}} \caption{ {\it Top panel}: Contour plot of ${\log}$ the stellar surface density (in units of $10^{10} M_\odot h/{\rm kpc}^2$) for g70q1o3 at $t=4.0\gy$. {\it Bottom panel}: Line of sight velocity contours of g70q1o3 at $t=4.0\gy$. In both panels the galaxy is seen edge-on. } \label{fig:r_map} 
\end{figure}

\subsection{Galactic model} \label{sec:galactic_model} We follow \cite{springel05} and model the dark matter (DM) halo as a \cite{hernquist90} profile 
\begin{equation}
	\label{eq:m_lh} \rho(r)=\frac{M_{dm}}{2\pi}\frac{a}{r(r+a)^3} 
\end{equation}
where $M_{dm}$ is the halo mass for $r\rightarrow\infty$, and $a$ is a scale length. The length scale $a$ can be matched to the concentration parameter $c$ and the scale $r_s$ in the Navarro, Frenk \& White density profile, $\rho_{_{\rm NFW}}(r) $ \cite[here after NFW]{nfw1,nfw2} by demanding that $\rho(r)=\rho_{_{\rm NFW}}(r)$ for $r\ll r_{200}$, where $r_{200}$ is the radius within which the mean DM density is 200 times the critical density of the universe. This yields the relation 
\begin{equation}
	\label{eq:a_lh_nfw} a=r_s\sqrt{2[{\rm ln}(1+c)-c/(1+c)]}\; . 
\end{equation}
Here we use for all galaxies $a=42\kpc$, $r_s=25\kpc$, and $c=9$.

We adopt an exponential stellar disk with a length scale $h$, and with vertical structure of an isothermal sheet with a constant height scale $z_0$ 
\begin{equation}
	\label{eq:m_disk} \rho_\star({\bf R},z)=\frac{M_\star}{4\pi zh^2}\;{\rm sech}^2 \left( \frac{z}{2z_0} \right)\; {\rm exp} \left( -\frac{R}{h}\right) 
\end{equation}
where ${\bf R}$ is the position vector projected in the disk plane. The gaseous component has the same radial structure as the stellar component, and its vertical structure is determined by the self-consistent solution of the hydrostatic equilibrium and Poison equations; We generate the galaxies with both collisional and collisionless components in equilibrium using a new method, as explained in appendix \ref{app:galaxy_generation}. 
\begin{figure}
	\centerline{\epsfig{figure=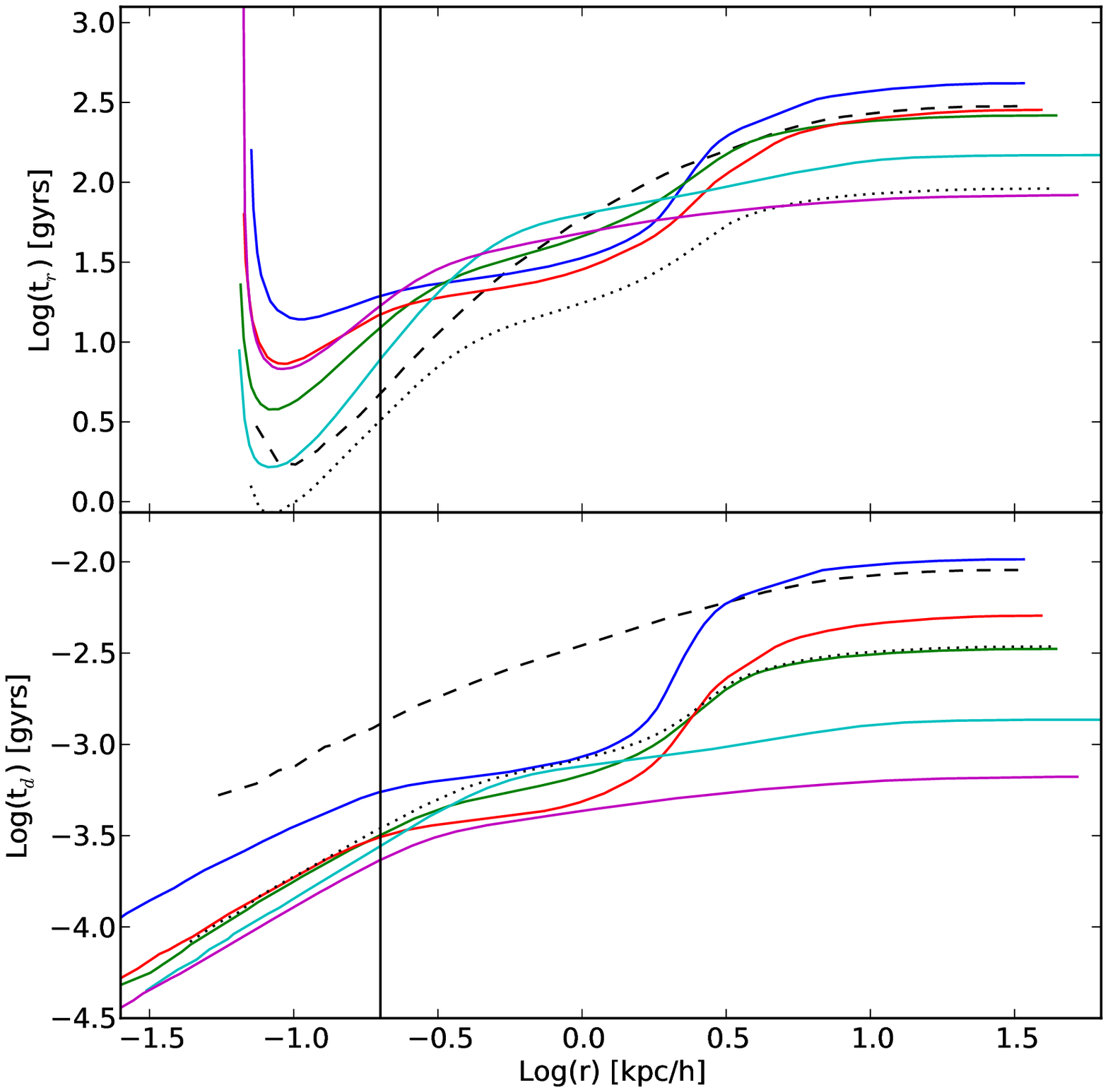, width=230pt, height=230pt}} \caption{ {\it Top panel}: Relaxation time of newly formed stellar components in at $t_f$, as function of distance from the center. Dashed black curve is for g30q1o3, dotted black for g70q1o3n2, blue for g30q025o3, red for g70q025o3, cyan for g70q1o1, purple for g70q1o2, and green for g70q1o3. The vertical solid black line denotes $r=4\epsilon_{sf}$, which we consider as the spatial resolution limit. {\it Bottom panel}: Dynamical time of newly formed stellar component in remnants at time $t_{rf}$, as function of distance of center. } \label{fig:r_rt_dt} 
\end{figure}
The used values of all relevant parameters are listed in table \ref{tab:galaxies} and explained in details in the caption. Note that some parameter choices, like the disk gas fractions, are similar to those used in previous work \citep{springel05b, robertson06}, and are essential to achieve disk-like remnants.

\section{The merging simulations} \label{sec:merging} We perform seven merging simulations, each of identical pairs of pure disk galaxies, as described in the previous section. We use three distinct orbits. Initially, galaxies are separated by $143\kpc$, each galactic disk in the merging pair is oriented in the same direction, defining the $y$ axis, and initial velocities are as summarised in table \ref{tab:mergers}. Initially all galaxies rotate in the same direction, and their angular momentum is negatively oriented in the $y$ axis. 

\begin{center}
	\begin{table*}
		\begin{tabular}
			{ c | c | c | c | c | c | c | c | c | c | c | c | c | c | c | c } \hline galaxy &$N_{dm}$&$N_\star$&$N_g$ &$N_{sph}$&$f_g$& $q$ & $a$ & $h$ &$z_0$ &$M_{dm}$ &$M_d$ &$\epsilon_{dm}$&$\epsilon_\star$&$\epsilon_{sf}$&$\epsilon_g$ \\
			name &$10^4$ &$10^4$ &$10^4$& & &- &$\kpc$&$\kpc$&$\kpc$&&&$\kpc$ &$\kpc$ &$\kpc$ &$\kpc$ \\
			\hline \hline g30q1 &16 &1.2 &2.8 &64 &0.3 & 1 &42 &6 &0.65 &200 &5.7 &0.75 &0.8 &0.07 &$min(\epsilon_{sph},0.07)$ \\
			\hline g30q025 &16 &1.2 &2.8 &64 &0.3 & 0.25&42 &6 &0.65 &200 &5.7 &0.75 &0.8 &0.07 & $min(\epsilon_{sph},0.07)$ \\
			\hline g70q1 &16 &1.2 &2.8 &64 &0.7 & 1 &42 &6 &0.65 &200 &5.7 &0.75 &0.8 &0.07 & $min(\epsilon_{sph},0.07)$ \\
			\hline g70q025 &16 &1.2 &2.8 &64 &0.7 & 0.25&42 &6 &0.65 &200 &5.7 &0.75 &0.8 &0.07 & $min(\epsilon_{sph},0.07)$ \\
			\hline g0 &160 &5 &0 &- &0.7 & - &42 &6 &0.65 &200 &5.7 &0.45 &0.3 &- &- \\
			\hline 
		\end{tabular}
		\caption{ Parameters used in the generation of the original galaxies. $N_{dm}$: number of DM particles. $N_\star$: number of stellar particles in the disk. $N_g$: number of gaseous particles in the disk. $N_{sph}$: number of neighbours used in the SPH calculation, $f_g$: gas mass fraction in the disk. $q$: pressurization parameter (has some effect on the effective equation of state, see \S \ref{sec:merging}). $a$: scale length of the DM halo. $h$: radial scale length of the disk. $z_0$: height of the disk. $M_{dm}$: DM mass (in $10^{10}\sm$). $M_d$: total (gas+stars) disk mass (in $10^{10}\sm$). $\epsilon_{dm}$: force softening length for the DM particles. $\epsilon_\star$: force softening for the initial stellar particles. $\epsilon_{sf}$: force softening for the newly formed stellar particles. $\epsilon_g$: gravitational force softening for the gas particles, equal to $min( \epsilon_{sph},0.07\kpc) $ where $\epsilon_{sph}$ is determined according to the nearest 64 particles. 	\label{tab:galaxies}} 
	\end{table*}
\end{center}
\begin{center}
	\begin{table*}
		\begin{tabular}
			{ c | c | c | c | c | c | c | c | c | c | c | l} \hline simulation &galaxies & $x$ & $y $ & $z$ & $v_x$ & $v_y$ & $v_z$ &$n_s$&$t_{rf}$&$t_{max}$&orbit \\
			name &merged &$\kpc$&$\kpc$&$\kpc$&$\kms$&$\kms$&$\kms$&- &$\gy$ &$\gy$&type \\
			\hline \hline g30q025o3 &g30q025 &0 &143 &0 &0 &-193 &-137 &6 &1.7 &1.8&polar \\
			\hline g30q1o3 &g30q1 &0 &143 &0 &0 &-193 &-137 &6 &- &1.8&polar \\
			\hline g70q025o3 &g70q025 &0 &143 &0 &0 &-193 &-137 &6 &1.7 &1.8&polar \\
			\hline g70q1o1 &g70q1 &143 &0 &0 &-193 &0 &+137 &6 &2.0 &4.0&both prograde \\
			\hline g70q1o2 &g70q1 &143 &0 &0 &-193 &0 &-137 &6 &2.0 &2.1&both retrograde \\
			\hline q70q1o3 &g70q1 &0 &143 &0 &0 &-193 &-137 &6 &2.0 &4.0&polar \\
			\hline g70q1o3n2 &g70q1 &0 &143 &0 &0 &-193 &-137 &2 &2.0 &4.0&polar \\
			\hline 
		\end{tabular}
		\caption{ Parameters used in the merging simulations. $x,y $ \& $z$: initial relative positions between the two merging galaxies. $v_x,v_y $ \& $v_z$: initial relative velocities. $n_s$: the maximum number of stellar particles that a gas particle may spawn. $t_{rf}$: is an approximate estimate for the formation time of the remnant. $t_{max}$: the time at which simulation was stopped.	\label{tab:mergers}} 
	\end{table*}
\end{center}

The galaxies begin to rotate around the center of mass, while tidal forces and dynamical friction bring them closer with time. When galaxies are close enough, their galactic disks collide for the first time, and a first starburst is triggered. Close encounters of the disks significantly enhance the gas density, setting ripe conditions for starbursts to occur. This behaviour is shown in Fig. \ref{fig:r_sfh}, where most simulations show a secondary starburst peak before a main one. Most of the stellar mass of the core of the remnant is produced in the main starburst corresponding to the final coalescence of the disks.

We denote the approximate time when the remnant is formed by $t_{rf}$, and loosely define it as the time when the core mass and velocity structure is no longer changing significantly during a dynamical time (see Fig. \ref{fig:r_eq}) We continue the simulations up to time $t_f$. These times are given in table \ref{tab:mergers}. Fig. \ref{fig:r_map} showcases a remnant at $t_f$. This remnant contains a thick disk made exclusively of newly formed stars. This behaviour has been seen in numerous other studies, e.g. \cite{springel05b, robertson06}.

In the analysis of the remnants, given below, we only consider separations larger than $4\epsilon$, where $\epsilon$ is the spline smoothing of the gravitational force. For this thresholdthe simulated remnants have a relaxation time longer than the simulated time, see Fig. \ref{fig:r_rt_dt} and the results are stable with time. Further, at this separation the force between two simulated particles is exactly Newtonian (in Gadget this happens at $2.5\epsilon$). Repeating the analysis with a threshold at $3\epsilon$, yields similar results. See \S \ref{sec:num_disc} for a detailed discussion of numerical considerations.

\section{Structure of the remnants} \label{sec:remnants}
The remnants contain a compact pseudo-bulge, a thick stellar disk, a large gaseous disk, and an extended diffuse stellar halo consisting of old stellar material originating from progenitor disks. This structure is typical of such simulations, see for e.g. \cite{springel05b, robertson06}. A thick stellar disk is showcased in Fig. \ref{fig:r_map}. according to Fig. \ref{fig:r_eq} (top panel) the stellar halo consist of $\sim 1.5 \times 10^{10} \sm$ in regions outside the stellar disk $6<r<60 \kpc$, and compared to the disk like components it is diffuse, and it is similar to known diffuse stellar haloes of disk galaxies in both mass and extent- the Milky Way stellar halo mass being $\sim 1 \times 10^9 \sm$ and extended up to $\sim 40 \kpc$ see for e.g. \cite{tumlinson10}.

Our analysis of the simulated remnants aims at mimicking the actual observations in as many details as possible. Thus, for each remnant, two-dimensional (2D) images are obtained projecting the three-dimensional (3D) particle positions and velocities onto 1000 planes corresponding, respectively, to random viewing directions limited by a maximum inclination of $\pi/6$ as to preserve a significant edge-on velocity component. On a random galaxy sample this means we only measure 1 out of 3 galaxies.

\begin{figure*}
	\centerline{\epsfig{figure=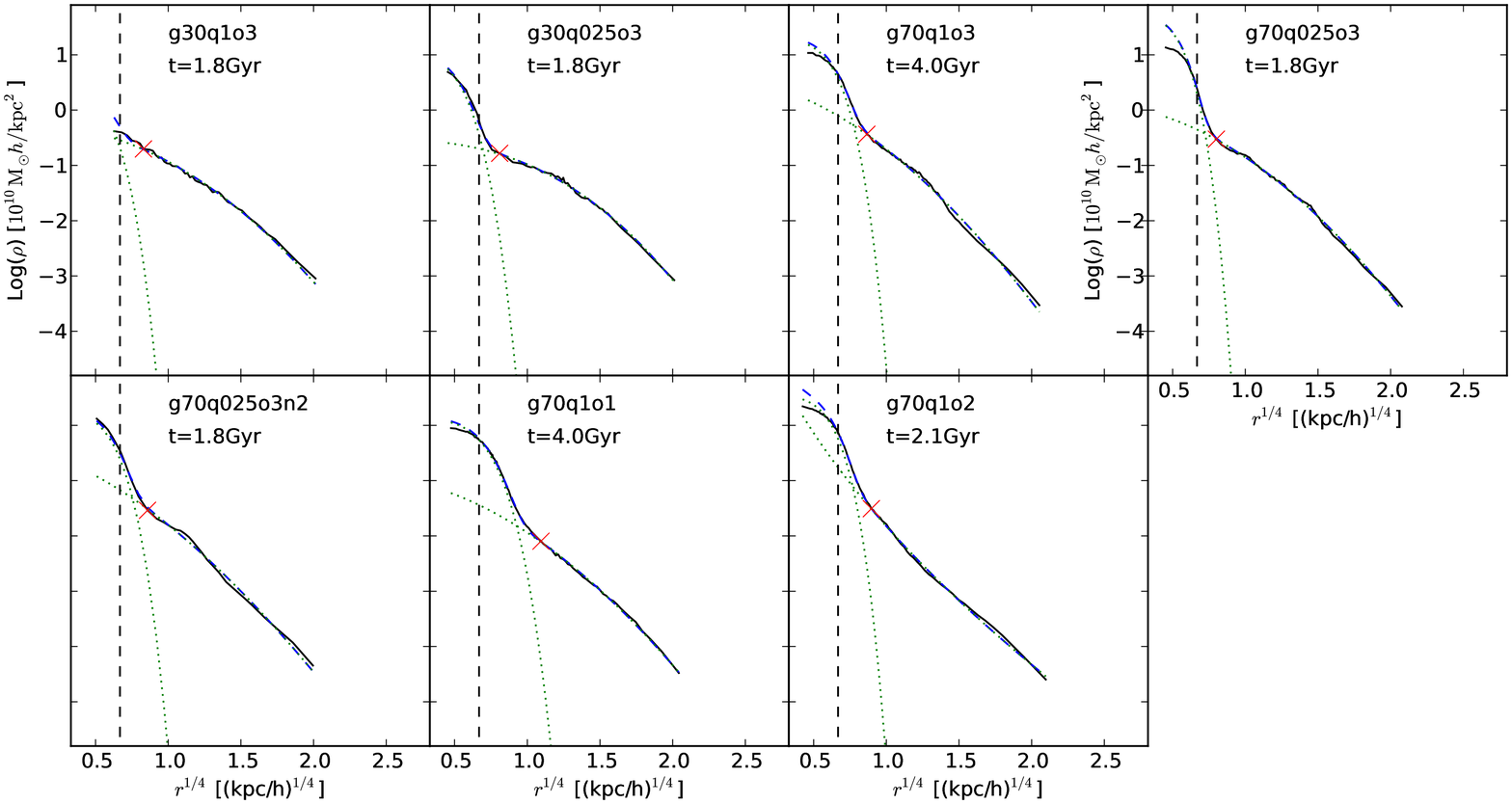, width=460pt, height=230pt}} \caption{Density profile of the remnants as a function of distance from the center. Green dotted lines are the S\`ersic fit. Vertical dashed lines denote formal numerical resolution, and red cross is $R_{max}$. } \label{fig:r_dens_prof} 
\end{figure*}

In order to determine the structural parameters $R_e$ and $n$, as appearing in Eq. \ref{eq:sersic}, we decompose the projected density of all stellar matter in the remnants as a sum of two circular S\`ersic functions. The fit is performed up to the formal simulation spatial resolution. A decomposition of remnants into two S\`ersic functions is shown in Fig. \ref{fig:r_dens_prof}.

\begin{figure}
	\centerline{\epsfig{figure=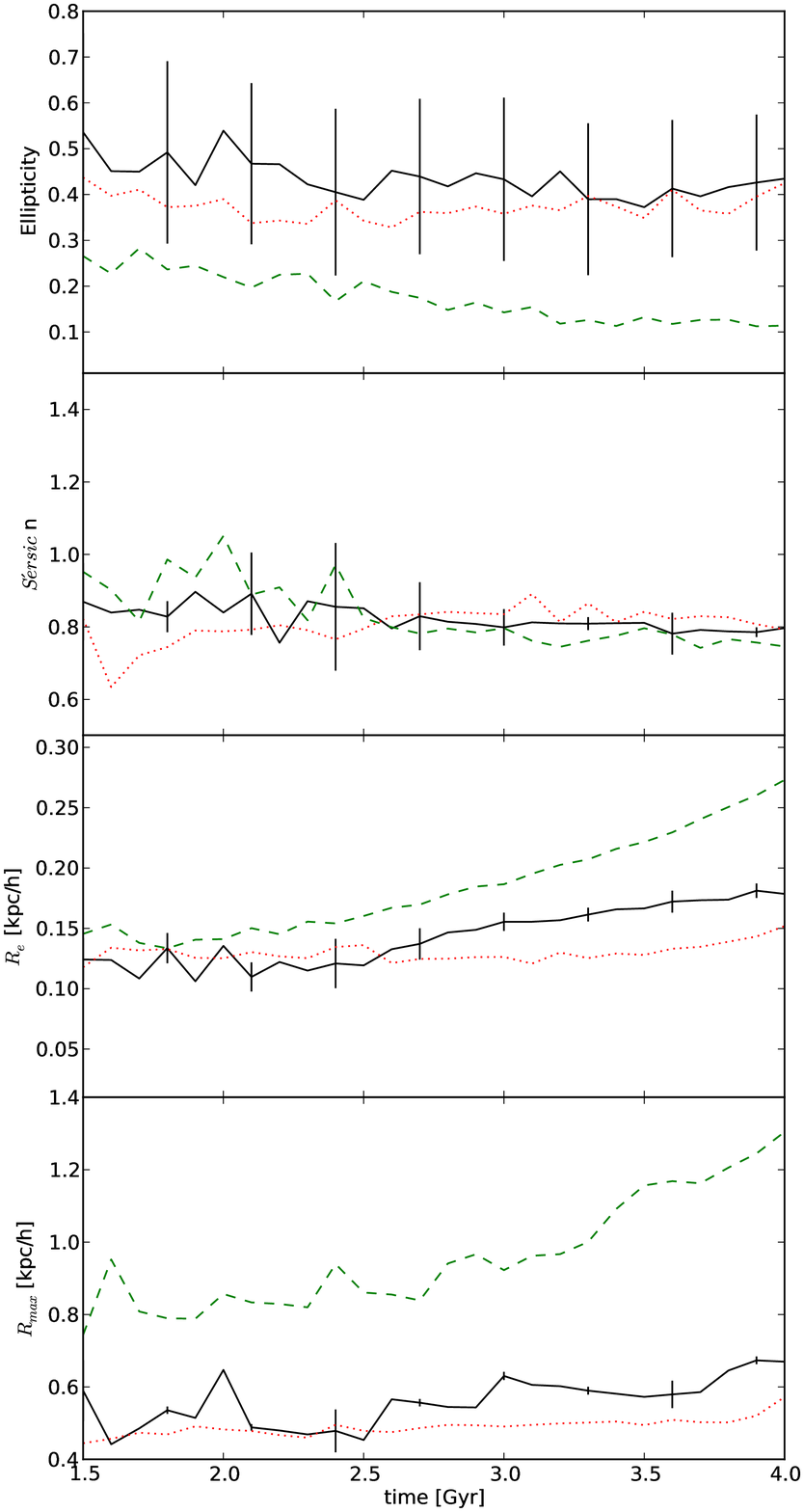, width=230pt, height=460pt}} \caption{S\`ersic parameters of the remnants. Dotted red, solid black, and green dashed curves, correspond to g70q1o3, g70q1o3n2, and g70q1o1, respectivly.} \label{fig:sersic} 
\end{figure}

We measure the ellipticity of the core. We do so by matching an ellipse to the isodensity contour containing most of the core mass. We define this mass as the projected mass within the circle with radius $R_{max}$, which in turn is defined as the circular radius where the core density is $5\%$ stronger than the outer component. This is a somewhat arbitrary choice, but we find results to be only weakly dependent on it. The results are shown in Fig. \ref{fig:sersic}. 

\begin{figure}
	\centerline{\epsfig{figure=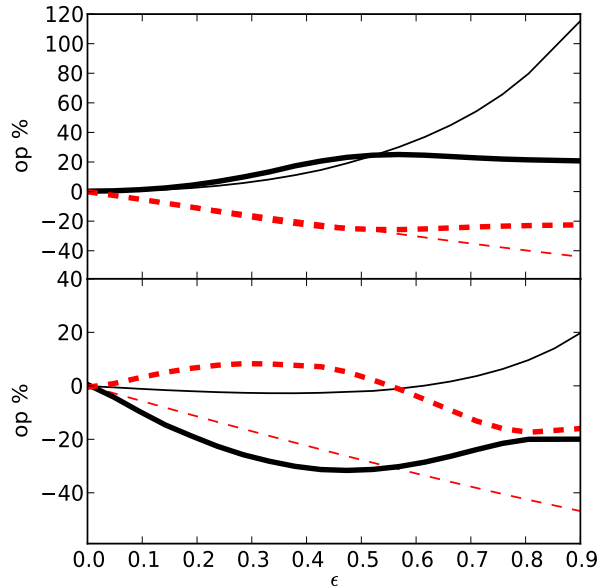, width=230pt, height=230pt}} \caption{Systematic over-predictions of S\`ersic parameters as a function of the remnants ellipticity, measured as described in the text. Dashed red and solid lines correspond to $R_e$, and $n$ respectively. Heavy and light lines are for $R_{min}=1.1\times R_e$, and $R_{min}=0$. Top panel is for $n=0.7$, and bottom for $n=3.5$.} \label{fig:serfit} 
\end{figure}

Observers usually fit images with a 2D decomposition, rather than a circular one. In such decomposition each component may be described by the following S\`ersic function: 
\begin{equation}
	\label{eq:sersic2d} \Sigma(r)=I_e \exp \left\lbrace -b_n \left[ \left( \frac{1}{R_e} \sqrt{x^2+((1-\epsilon) y)^2} \right)^{1/n}-1 \right] \right\rbrace, 
\end{equation}
where the main remnant axis is defined as the $x$ axis. This decomposition gives different S\`ersic parameters when compared to the circular one. We calculate the S\`ersic parameters of the two dimensional decomposition using the following method: we find the structural parameters of the observed image, the ellipticity $\epsilon$, and the circular S\`ersic $R_e^{1D}$, and $n^{1D}$. Next we construct a $2D$ particle realization using the previously measured ellipticity $\epsilon$, and random two-dimensional S\`ersic parameters $R_e^{2D}$, and $n^{2D}$. The purpose is to find such $2D$ parameters that will reproduce the observed image. Thus we build many such $2D$ particle realizations, until we find the one that reproduces the circular $R_e^{1D}$, and $n^{1D}$. Two-dimensional parameters of the simulated remnants are given in table \ref{tab:self}. In general, we find that $n^{2D}$, for $n^{2D} \sim 1$, is usually larger than $n^{1D}$ for the same galaxy. This can be seen in Fig. \ref{fig:serfit}. In that figure, the over-prediction of a parameter, say $n$, is defined as 
\begin{equation}
	\label{eq:over_prediction} op\%=\frac{n^{1D}-n^{2D}}{n^{2D}}\times 100\% 
\end{equation}
where $n_c$, and $n_{2D}$ are the circular and the two-dimensional S\`ersic parameters of the remnants.

The most important structural feature of the remnants is the small S\`ersic index, $n<1$, of the core. Note that the core created in simulation g30q1o3 is very small, and is not dominant over the outer galactic component within the simulation spatial resolution. Thus, we do not include this core in most of the structural analyses. 
\begin{center}
	\begin{table}
		\begin{tabular}
			{ | c | c | c | c | c } \hline simulation &$e$ &$R_e$ &$n$ & $C/T$ \\
			name &- &$\kpc$&- & -\\
			\hline \hline g70q1o1 &0.1 &0.4 &0.79 & 0.41 \\
			\hline q70q1o3 &0.4 &0.29 &0.57 & 0.25 \\
			\hline g70q1o3n2 &0.45&0.37 &0.33 & 0.27 \\
			\hline g70q1o2 &0.4 &0.26 &0.67 & 0.42 \\
			\hline g70q025o3 &0.14&0.11 &0.85 & 0.1 \\
			\hline g30q025o3 &0.33&0.1 &0.79 & 0.03 \\
			\hline 
		\end{tabular}
		\caption{Two-dimensional S\`ersic parameters of the remnants core at $t_f$. The last column give the core to total light ratio.	\label{tab:self}} 
	\end{table}
\end{center}

\begin{figure*}
	\centerline{\epsfig{figure=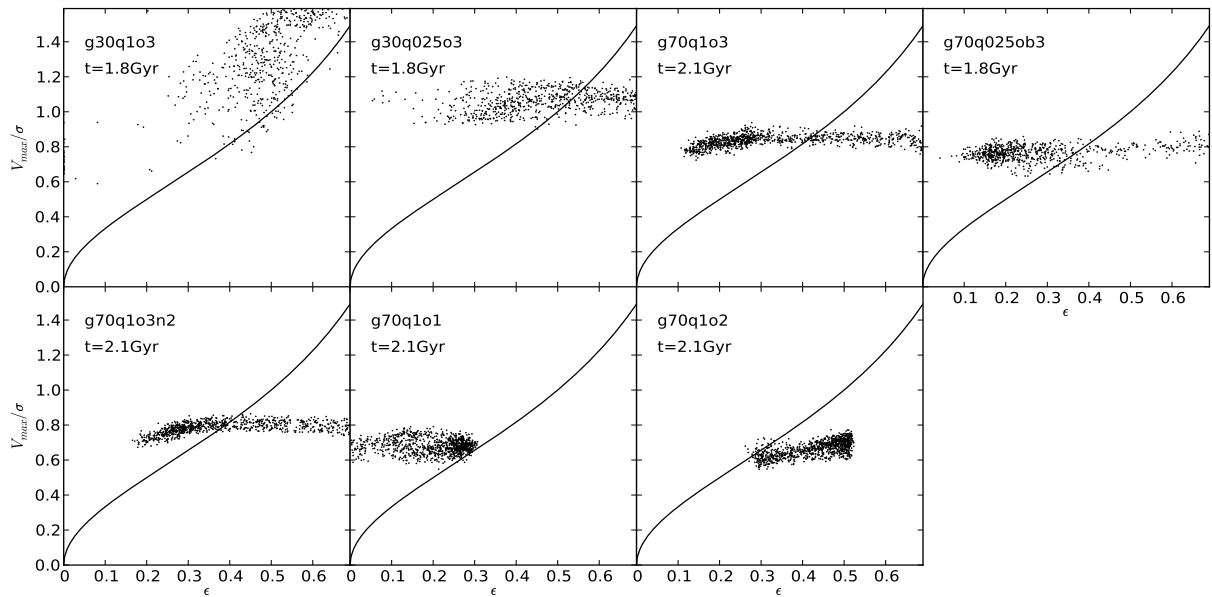, width=460pt, height=230pt}} \caption{Anisotropy diagram of remnants core. Each galaxy is viewed from 1000 random angles, each viewing angle is represented by a point.} \label{fig:r_n_iso} 
\end{figure*}

We quantify the kinematic structure of the remnants by positioning them in the anisotropy diagram \citep{binney87}. A point in this diagram consists of the ratio $\sigma/V_{max}$ versus the ellipticity $\epsilon$, where $V_{max}$ is the maximal rotational velocity of the central component. The information given by such a point may be interpreted as a measure of how much the gravitational system under inspection is supported by rotation. The solid lines in Fig. \ref{fig:r_n_iso}, under some simplifying assumptions, are expected values of $\sigma/V_{max}$ for highly rotationally supported systems with ellipticity $\epsilon$. Systems above this line are considered to be rotationally supported, and systems below it- supported by random motions. In Fig. \ref{fig:r_n_iso} we plot the different simulated remnants as viewed from $1000$ different angles, limited by inclination as described before. This plot shows that almost all simulated remnants appear rotationally supported for a significant fraction of viewing angles. This is true at least at a certain period of time in the galactic evolution.

Fig. \ref{fig:r_n_iso} is plotted using an observationally inspired method: Velocity is measured along a thin slit aligned with the remnant image main axis. The slit ratio is $12:1$, and length $2R_{max}$. It is divided into bins of $3000$ particles. We measure $V_{max}$ as the maximal velocity along the slit, and $\sigma$ as the velocity dispersion in the inner square of side length $R_{max}/6$. It is clear that a significant fraction of the viewing angles leads to a classification of the core as pseudo-bulge. As further analyses, we plot the fraction of viewing angles that leads to such classification as function of time. See Fig. \ref{fig:r_n_iso_per}. Even for our worst-case merger, and after $4\gy$ from the start of the simulation, more than $20\%$ of the viewing angles lead to a pseudo-bulge classification. 

\section{Numerical considerations} \label{sec:num_disc} In Fig. \ref{fig:r_eq} we show that the newly formed cores are indeed close to equilibrium, with the velocity moments in the observed range \citep{kormendy82}. Here we further investigate possible numerical effects related to the finite relaxation time and smoothing. The dynamical time scales of the remnants at their formation times $t_{rf}$ are shown in Fig. \ref{fig:r_rt_dt}. The dynamical time at distance $r$ from the center is calculated as $t_d(r)=R_{_{1/2}}/\sigma_{_{1/2}}$, where $R_{_{1/2}}$ is the radius enclosing half the mass within $r$, and $\sigma_{_{1/2}}$ is the velocity dispersion within $R_{_{1/2}}$. The remnants core relaxation time at time $t_f$ and near the center (up to the resolution limit) is longer than the simulated time. The relaxation time is calculated as \citep{farouki94} 
\begin{equation}
	\label{eq:tr} t_r=F\frac{N}{26\, {\rm log(\Lambda)}}t_d 
\end{equation}
where $N$ is the number of particles within $r$, $F$ is an empirical dimensionless factor, $t_d$ is the dynamical time at $r$, and $\Lambda=R_{_{1/2}}/\epsilon$. We use $F=8$.

Two Numerical considerations in the simulations presented in this paper may be considered as marginal; First, the relaxation time. Fig. \ref{fig:r_rt_dt} and Table \ref{tab:mergers} show the relaxation time of the remnants at the final time of the simulation, $t_f$. At the innermost studied region $r=4\epsilon$, the relaxation time, $t_r$, is always larger than the simulated time $t_f$. While $t_r \sim 3t_f$ for simulation g70q025o3, it is close to the simulated time for g30q1o3, and it is in between these values for the other simulations. Of course $t_f$ is longer than the age of the remnant which formed about $1\gy$ after the beginning of the simulation. Second, we analyse results in a region $1.6$ times larger than the distance at which gravitational force in the simulation becomes exactly Newtonian. 

\begin{figure}
	\centerline{\epsfig{figure=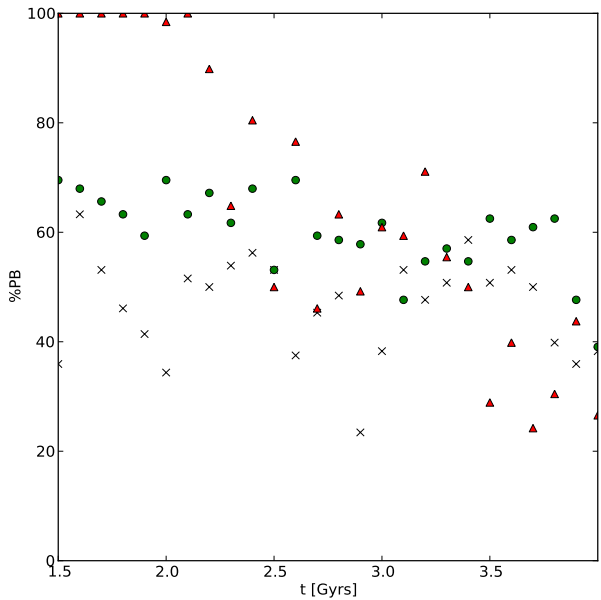, width=230pt, height=230pt}} \caption{Percentage of solid angles giving a rotationally supported remnant, as function of time. Black crosses, close green circles, and closed red triangles correspond to g70q1o3n2, g70q1o3, and g70q1o1, respectively.} \label{fig:r_n_iso_per} 
\end{figure}

We now discuss in more detail the above considerations. The gravothermal catastrophe \citep{lynden68} is one of the most important gravitational effects driven by collisions, either physical or numerical. Due to the negative heat capacity of such systems, if the center is hotter than the outskirts, heat flows outwords resulting in further heating and contraction of the central region. As the core collapses, it also rotates faster even though angular momentum flows outwards \citep{lynden74, hachisu79}.     The magnitude of these effects after a single relaxation time is modest, as can be seen in Fig. 4 in \cite{ernst07}. \cite{spurzem03} studied the evolution of the ratio $v_{rot}/\sigma$ and found that it monotonically decreases with time. Although this \cite{spurzem03} result seems favourable to ours, in order to classify a bulge as pseudo we need to know its ellipticity. Fig. 7a in \cite{einsel99} shows how the dynamical ellipticity $e_d=\sqrt{1-(1-e)^2}$ evolves with time. The evolution of the cores with maximal mean ellipticity (i.e. ellipticity averaged over viewing angles), $e \sim 0.5$, within a single relaxation time, is modest. To further assess the two-body relaxation effects on our results we use compare simulation g70q1o3n2 and g70q1o3. These two simulations are identical in every respect, except g70q1o3n2 contains a third of the newly formed stellar particles. Fig. \ref{fig:r_n_iso_per} shows that the fraction of viewing angles leading to a pseudo-bulge classification of the core of simulation g70q1o3n2 is lower than that in simulation g70q1o3. Thus we conclude that two-body relaxation effects, when using the anisotropy diagram, are more likely to lead to classical bulges rather than pseudo. Hence our identification of pseudo-bulges is robust.

Two numerical effects, the gravitational force smoothing, and core collapse due to relaxation, compete to decrease and increase, respectively, S\`ersic $n$. Under these circumstances, we would like to understand whether or not the density profile can be considered as physically converged in the region $r<4\epsilon$. Unfortunately this is a grey area, as for our case there is no good agreement between different studies: \cite{moore98} claim that the minimum resolved scales should be no larger than half the mean interparticle separation within the virial radius, and that several thousands of particles are needed to resolve the innermost regions of collisionless systems. These criteria are all met in the simulations here. However, \cite{ghigna00} claim that, in addition to a large number of particles, convergence requires that, in addition to a large number of particles, the minimal resolved radius is at least as large as three times the distance at which gravitational forces between two particles becomes Newtonian. This means we should have used at least $r>7\epsilon$, but these regions already miss some of the most interesting features of the bulge. However, we did not see any difference when we analysed the results with $r>3\epsilon$ compared to $r>4\epsilon$. The study of \cite{klypin01} claims that regions containing ~$200$ particles can already be reliably resolved provided $r>4\epsilon$. In \cite{power03} a new condition for convergence is presented. This condition is based on a maximum characteristic acceleration imprinted by softening. It is empirically studied for NFW haloes, and it is not clear how it is applicable to a galactic bulge.

As a final note, the bulges presented in this study are too bright when compared to observed bulges. They are outliers (from above) in the Kormendy relation of bulge brightness distribution \citep{kormendy77}. It is known that a realistic galaxy merging requires some sort of feedback (e.g. galactic winds) to prevent star formation runaway in the central regions. We did perform preliminary simulations including such winds, and find that simulated bulges obey the Kormendy relation. The problem is that these simulated bulges have less particles in their inner regions causing a dynamical relaxation time which is too short for our needs.

\section{discussion} \label{sec:discussion} The standard $\Lambda$CDM model naturally produces massive classical bulges in large spiral galaxies \citep{scana9, scana10}. This is likely the result of the significant late violent assembly of galactic material in $\Lambda$CDM \citep{peebles10}. Thus $\Lambda$CDM faces two problems when compared to observations of nearby galaxies. First, there are too many giant pure disk galaxies with none or tiny bulges (even pseudo-bulges). Second, there are too many large pseudo-bulges that are rotationally supported, in contrast to the tendency of major mergers to produce classical bulges with little rotational support. Here we have demonstrated that the existence of pure disk galaxies could be accompanied by the existence of large pseudo-bulges as natural outcome of mergers of pure disks. 

In contrast, the formation of pure disk dwarf galaxies could be explained by Supernovae (SN) feedback. SN feedback ejects gas from the central region, releasing most central stars from the gravitation grip in the already shallow gravitational potential wells of these small galaxies \citep{governato10, navarro96}. The formation of large pure disks within the $\Lambda$CDM model requires significantly more contrived conditions. Although our simulations are done with large original disk galaxies, we show the formation of pseudo-bulges only requires the existence of gas rich dwarf disk galaxies.

Gas rich last major mergers may produce the observed abundance of pseudo-bulges in the local Universe. Pseudo-bulges initially created by a major merger event may continue to grow over long periods of time by regular secular processes. In this picture pseudo-bulges are expected to contain stellar populations younger than classical bulges - not only due to the later secular evolution, but also because in this scenario a significant fraction of classical bulges is expected to form by merging of pseudo-bulges. This expectation is consistent with observations \citep{kormendy04}. Further, according to this picture the fraction of classical bulges should increase with the density of the environment, because the galactic merging rate depends on environment density; Higher fractions of mergers of pseudo-bulges into into bulges is expected in denser regions. The effect seems to be consistent with current observations \citep{kormendy09} but further evidence is required.

As a final note, \cite{kormendy2011} confirm the correlation between black-hole (BH) mass and their host classical bulges. However they find that BH masses correlate little with pseudo-bulges and not at all with their host disks. The common interpretation these results is that BHs co-evolve with their host classical bulge, but not with their host disk. Continuing this line of thought, the finding that BH mass may correlate with its host pseudo-bulge but not with its host disk means that some pseudo-bulges may not evolve from disks by secular evolution. The pseudo-bulge formation model presented in this work may supply the mechanism needed to form such pseudo-bulges: pseudo-bulges that did not evolve from their host disk.

All codes used in this paper (except for the modified version of Gadget2) are available at \\* \textcolor{blue}{\underline{\url{http://physics.technion.ac.il/~kari/}}}

\section*{Acknowledgments} We would like to thank Prof. John Kormendy for his useful comments regarding the definition of the problem of bulgeless galaxies, and Prof. Volker Springel for giving us access to the modified version of Gadget2.

\bibliographystyle{mn2e} 
\bibliography{KN11} \bsp

\appendix \section{The reconstruction of isolated galaxies in equilibrium} \label{app:galaxy_generation} The simulated galaxies contain both a collision-less (DM and stars) and a collisional (gas) component. We initially build a galaxy in detailed equilibrium, in which each component is in kinematic equilibrium regardless of the others. We describe the method used for the collision-less components first, and then the method used for the gaseous component.

\begin{figure}
	\centerline{\epsfig{figure=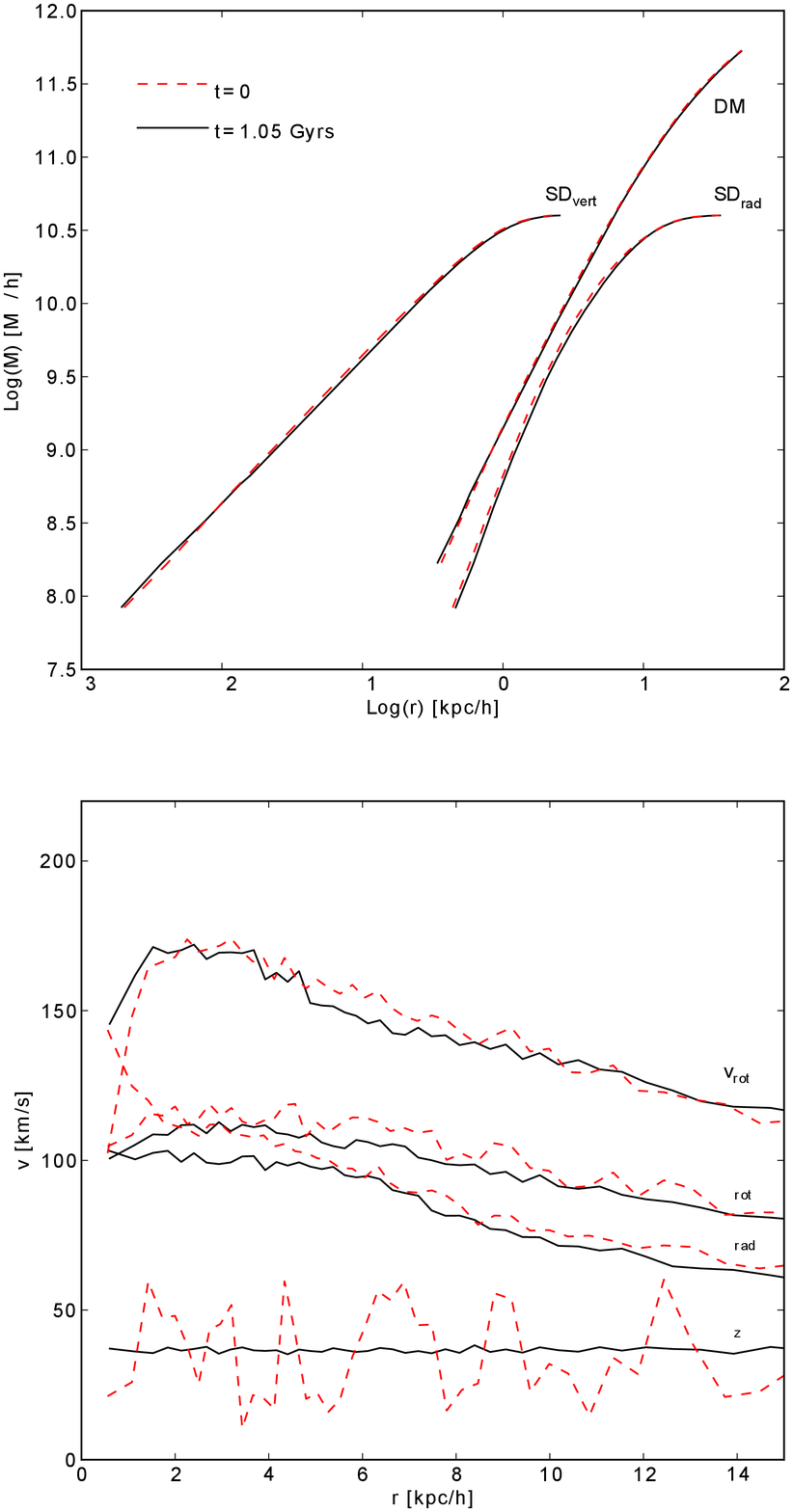, width=230pt, height=460pt}} \caption{Equilibrium of the DM and the stellar disk (SD) of the gas-less galaxy $g0$. Mass distribution (top) and rotational velocity and velocity dispersions of the stellar disk (bottom) shown at $t=0$ and $t=1.05\gy$ corresponding to $11t_{dyn}$ where the dynamical time $t_{dyn}$ is computed at radius $r=6\kpc$. The curve $SD_{rad}$ is the mass of the disk contained in a sphere of radius r. The curve $SD_{vert}$ shows the stellar mass contained between two planes at distance $r$ above and below the mid-plane of the disk. } \label{fig:g_ng} 
\end{figure}

\begin{figure*}
	\centerline{\epsfig{figure=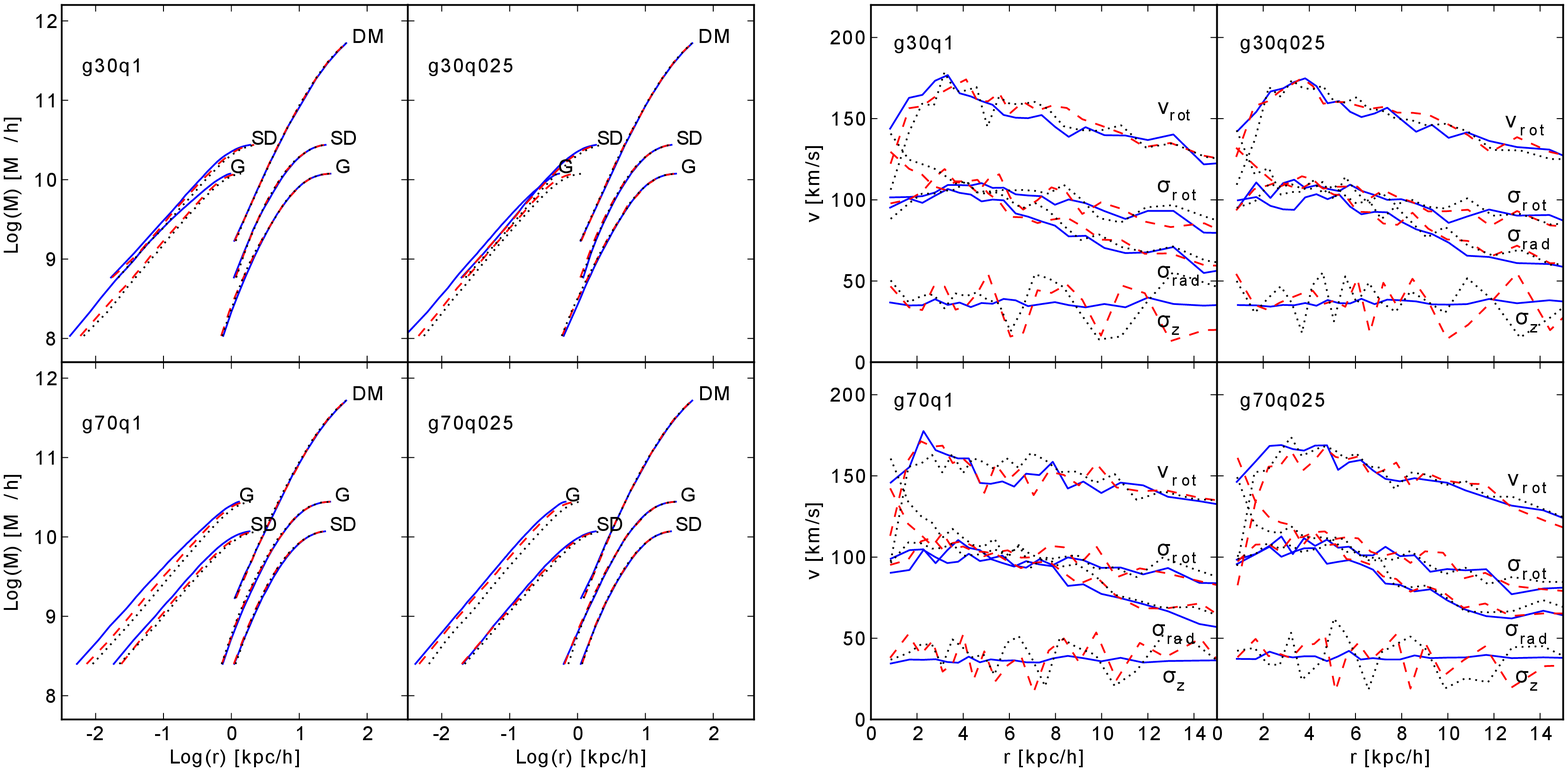, width=460pt, height=230pt}} \caption{ {\it Left panel}: Equilibrium mass profiles for the DM, stellar disks (SD) and gas (G). In each panel, the curves to the left show the mass enclosed between two planes parallel to the stellar disk, enclosing it, and each at a distance $r$ from it. Solid blue, dashed red and dotted black lines, respectively correspond to $t=0$, $3t_{dyn}$, and $t=6t_{dyn}$. {\it Right panel}: Velocity moments equilibrium of stellar component, corresponding to the SD curves in the left panel. The notation of the curves is the same as in the left panel. } \label{fig:g_eq} 
\end{figure*}
Equilibrium initial conditions for N-body simulations are solutions to the collisionless Boltzmann equation. There are numerous methods to obtain particle positions and velocities given some spatial or phase restrictions \citep[e.g.][]{sch79,kuij95,wid05,hern93}. most methods have some severe restrictions, and are restricted to either a single-component, some special symmetry, collisionless components (do not allow for inclusion of a heavy gaseous component), or are merely approximate. We now describe a method which is efficient, general, and accurate. We develop this method mainly as a derivation of the iterative method (IM), which was introduced by \cite{rodionov09}. This method belongs to the general family of ``Made-to-Measure" (M2M) methods first introduced by \cite{syer96}. The basic algorithm of the IM method is: 
\begin{enumerate}
	\item Generate a particle realization of all components of the galaxy. This realization should include only particle positions, while velocities are not essential, although certain physically motivated choices would boost convergence. We call this realization the ``original" snapshot. 
	\item \label{itm:tint}For each galactic component, use an N-body code to move the particles forward to a new snapshot. The integration is done for a dynamically significant period of time, $\tau$. All other components are unaltered and their gravitational potentials are assumed static.
	
	\item Transfer velocities from the new to original snapshot. Assign each particle an integer number $n_u$ equals the number of times the particle has been used for velocity transfer. For every particle in the original snapshot, locate the closest $n_p$ particles in the new snapshot. Among those particles, find the nearest one with the smallest $n_u$. Transfer the velocity of this particle to the original particle. This is done separately for every galactic component. 
	\item repeat from step \ref{itm:tint}, until sufficient convergence is achieved. 
\end{enumerate}
This method slowly assigns a developed kinematic state which corresponds to a given matter density. This method is also very flexible, and easy to implement. The kinematics may be constrained in many ways (e.g. isotropic or rotating bulges). This method has many advantages over approximate ones to produce multi-component gravitating systems in equilibrium; For example, methods using a Gaussian velocity distribution with dispersion derived from the Jeans equations \citep[e.g.][]{springel05}, may result in components that are out of equilibrium in their central regions.

\cite{dehnen09} developed a modified M2M method (MM2M) in order to alleviate two main problems of the IM method. The first is a scale problem- for every galactic component, the regions which are close to the galactic centre have a much shorter dynamical time than the outer regions. When simulating with an N-body method in the IM, the inner regions achieve sufficient convergence much faster than the outer regions, yet they are further simulated for a long time until the whole model, including outer regions, achieve sufficient convergence. Second, The method requires a full N-Body simulation with time-cost comparable to the simulation itself; i.e., the method requires a computational effort similar to that of the simulation. These problems are not straight forward to solve; \cite{rodionov08} showed that using a static potential instead of integrating with the N-Body method results in unprovable and non-physical solutions to the Boltzmann equation, such as oscillating velocity profiles.

We now describe the method used in this work, which we term the ``Modified Iterative Method" (MIM). This method has equivalent benefits to MM2M, albeit it is achieved with a simpler technique, and is also used and tested with massive gaseous disks. The algorithm is as follows: 
\begin{enumerate}
	\item \label{itm:gen} As in the IM method, generate a particle realization of all components of the galaxy 
	\item Integrate in time all particle trajectories in the global static potential. Integration time $\tau_i=n_i t_{di}$ is individual per-particle, with $t_{di}$ being some local dynamical time, and $n_i$ some small integer number common to all particles. 
	\item Transfer velocities from new to original snapshot: this is done like in the IM method, except copied velocities are given a ``penalty" in form of addition of some random velocity proportional to the number of times it was used, $n_u$. 
	\item repeat from step \ref{itm:gen}, until sufficient convergence is achieved. 
\end{enumerate}
The changes purpose are to better explore the phase-space to achieve better and faster conversion. With these changes, gravitational systems converge to a physically probable Boltzmann solution even when integrating orbits using a static potential.

In order to include a massive gaseous disk in equilibrium, we integrate the collisionless particle trajectories under the static potential of a gaseous disk. While the radial structure of this disk is arbitrarily chosen, the vertical structure is a self consistent solution to the hydrodynamical and Poisson equations. To achieve this, we modified the N-Body code to allow the gaseous particle to move only in the vertical direction. In this study, gas particles are initially positioned as a realization of eq. \ref{eq:m_disk}. Applying a dissipative term to their velocity, and applying the static potential from the collisionless components, in addition to their self-gravity, the particles rapidly sink to equilibrium. We find that convergence can be accelerated if using a vertical scale factor $z_0$ that is much smaller than the equilibrium one. This is because the driving force in this configuration is the pressure, which in this case is much stronger than gravitational.

This method allows for fast construction of galaxies with massive gaseous disks, is arguably simpler to implement compared to methods with similar purpose, and more importantly, gives an exact solution to the Boltzmann equation. This method is very flexible, can be used with arbitrary spheroidal components, and requires no analytical knowledge of their kinematic properties.

We test the ability of the MIM at producing equilibrium disk galaxies. Fig. \ref{fig:g_ng} shows the results of simulating the gas free galaxy g0 in isolation for 11 dynamical times at a nominal radius of $6\kpc$, where $t_d=0.15\gy$. Fig. \ref{fig:g_ng} shows that there is no significant change in mass structure, or velocity structure, over this time period. In order to test the performance of the method when a gaseous disk is included, we run the additional 4 original galaxies for $6$ dynamical times, as measured at our nominal radius of $6\kpc$. The results are shown in Fig. \ref{fig:g_eq}, where the solid blue, dotted black, and dashed red curves correspond to $t=0$, $t=3t_{dyn}$, and $t=6t_{dyn}$. When comparing to the g0 simulation, it is evident that the massive gaseous disk damages the quality of equilibrium achieved with the MIM method. The reason behind this is that gas instabilities change the structure of the gaseous disk (as seen by its the changing thickness). However, equilibrium is reached within a short time, as seen by the small difference between the dashed red lines and the dotted black lines. This result applies both for mass and velocity structure of the galaxies, as shown in the right panel of the same figure. In the merging simulations the galaxies are initially separated far enough as to allow the gaseous disks to achieve better equilibrium. 

\label{lastpage} \end{document}